\documentclass[letterpaper]{JHEP}
\usepackage[dvips]{epsfig}
\usepackage{epsfig}
\usepackage{graphicx}

\title{Macroscopic open strings and gravitational confinement}
\author{Ruth Gregory\\ Centre for Particle Theory, 
University of Durham,\\ South Road, Durham, DH1 3LE, U.K.}

\abstract{
We consider classical solutions for strings ending on magnetically
charged black holes in four-dimensional Kaluza-Klein theory. We 
examine the classical superstring and the global vortex, which can
be viewed as a nonsingular model for the superstring. We show how 
both of these can end on a Kaluza-Klein monopole in the absence of 
self-gravity. Including gravitational back-reaction gives rise to
a confinement mechanism of the magnetic flux of the black hole along
the direction of the string. We discuss the relation of this work 
to localized solutions in ten dimensional supergravity.
}

\preprint{DCPT/02/61, hep-th/0208100}

\def\real{I\negthinspace R}

\def\half{\textstyle{1\over2}}

\def\quarter{\textstyle{1\over4}}

\def\ie{{\it i.e.,}}

\newcommand{\be}{\begin{equation}}
\newcommand{\ee}{\end{equation}}
\newcommand{\bea}{\begin{eqnarray}}
\newcommand{\eea}{\end{eqnarray}}
\newcommand{\bml}{\begin{mathletters}}
\newcommand{\eml}{\end{mathletters}}

\begin{document}

\section{Introduction}

The interplay between macroscopic soliton solutions in field theory,
and microscopic quantum physics has always been a fruitful and fascinating
one. Solitons, while generally `heavy' and `classical' in nature, 
nonetheless are an important and indeed integral 
part of the particle spectrum of the theory. The effort to fully
integrate these objects into quantum theory has led to a deeper 
understanding of existing theories. In particular the study of 
the `solitonic' D-branes \cite{DBr} has revolutionized the
study of string theory.

There are two different perspectives from which one can view
a D-brane; one is an inherently stringy point of view,
which is that of applying Dirichlet, rather 
than Neumann, boundary conditions to open strings \cite{Dirich}. 
On the other hand, a somewhat more `classical'  perspective views 
the brane as a possibly extended solution of low energy string supergravity 
carrying Ramond-Ramond (RR) charge \cite{SUsol}. 
Combining these two approaches has led to many new insights, such as 
a better understanding of black hole entropy \cite{ENT} and the adS/CFT
correspondence \cite{ADS}.

There are also classical solutions carrying Neveu-Schwarz
charge, the NS5-brane and the string. This latter solution is often 
viewed as a macroscopic fundamental superstring \cite{MAC}. 
Viewing strings and D-branes classically as black brane solutions 
to supergravity, one can ask a whole range of interesting gravitational 
questions, such as what happens when two branes meet? 
There are many known solutions for intersecting branes (see \cite{Jerome} 
and references therein), however these often `de-localized' in the 
sense that the solution depends only on the mutually orthogonal directions.
A genuinely localized intersection has proved somewhat elusive,
although one can construct near-horizon solutions (e.g.\ \cite{NH}),
and semi-localized solutions \cite{LINT}, indeed, 
in some cases localization is thought not to be possible \cite{bald}.  

If however this rather nice
correspondence between supergravity solutions and the
stringy picture is neat and closed, then there surely should exist
solutions of fundamental strings terminating on D-branes.
This statement has some secondary
implications. If it is indeed possible to find a classical solution for
a string terminating on a D-brane, then it should in principle
be possible for a macroscopic fundamental string to split, in a manner 
similar to the four-dimensional cosmic string \cite{split, smooth}, 
since it can nucleate a D/anti D-brane
pair along its length. This could have potential implications
for the dynamics of a superstring network in the early universe, 
where additional interactions
of string splitting and rejoining would also have to be taken into account.
 
To explore this question we consider a rather simpler one. Rather
than examining the ten-dimensional problem, for which one might expect
that fully localised intersecting brane solutions would be a 
necessary precursor, we look at two rather simpler toy models: a classical
superstring, and its nonsingular field-theoretic cousin -- the global 
vortex -- in four-dimensional Kaluza-Klein (KK) gravity. 
This `ministring' can be viewed either as a truncation of
the IIA superstring, or the dimensional reduction of a membrane from
five-dimensional gravity \cite{AETW}.
These strings can split by the nucleation of the four-dimensional 
equivalent of D0 or D6 branes, which are now the electric and 
magnetic KK black holes \cite{KKBH}.
The D0 brane is a gravitational wave travelling around the internal
circle, and is singular, however, the `D6' brane is the KK
monopole\cite{KKM}, and is constructed from the Taub-NUT instanton, 
which, at least for unit monopole charge, is completely regular from
the five-dimensional point of view. 
 
In what follows we start by taking a rather empirical approach, initially 
ignoring the gravitational back reaction of the string (working in the limit
in which the string is arbitrarily light compared to the monopole) first
exhibiting a string worldsheet which satisfies charge conservation
in the sense of the NSNS and RR form fields that live in the spacetime. 
Similarly, for the global string, initially, we solve only the field equations
on the KK monopole background, before coupling in gravity.
Even this background or `probe' solution already gives important clues about
the gravitational backreaction.

\section{Background solutions}

In order to find a configuration corresponding to a classical superstring
ending on a monopole, consider its interpretation as a double-dimensional
reduction of a supermembrane. We are then looking for a configuration
involving a  membrane in five dimensions terminating on the core of the 
KK monopole which is nonsingular from the five dimensional point of view,
therefore the membrane cannot have any ends. Fortunately, because of
the Hopf-fibration of the KK monopole, it is 
easy to see how to wrap the membrane so that it has no ends, 
yet appears to terminate from the dimensionally reduced point of view: 
one wraps the membrane around the fifth dimension along 
the `south' direction of the Taub-NUT
instanton, but not along the `north' direction. 
This is a regular configuration
with no boundaries (other than that at infinity) and yet from the four
dimensional point of view looks like a string with an end. 
This problem has in fact been well studied in 
the M-theory context \cite{BGr,wrap}, where there are families
of static M2-branes given by holomorphic curves on the (hyper-K\"ahler)
Taub-NUT manifold. These have the lower dimensional interpretation of
D2 branes in the background of a D6 brane with fundamental string charge.
The configuration we have is a five dimensional version of the
limit of one of the curves
presented in \cite{BGr} in which a D2-brane is dragged past the 
KK-monopole leaving a fundamental string 
connecting the core of the monopole to infinity.
 
This clearly represents a geometric solution to the problem, but can it be
consistently coupled in to gravity and the 3-form field that also exists
in the toy 5-dimensional gravity? 
To answer this, consider a nonsingular field-theoretic model 
of the classical superstring -- the global vortex -- which is a 
topological defect solution to the U(1) theory defined by the lagrangian
\bea
{\cal L} &=& |\nabla_\mu\Phi|^2 - {\lambda\over 4} (|\Phi|^2 - \eta^2)^2
\nonumber \\
&=& \eta^2 \left \{ (\nabla_\mu X)^2 + X^2 (\nabla_\mu \chi)^2 -
{\lambda\eta^2\over4} (X^2-1)^2 \right \}
\eea
writing $\Phi = \eta X e^{i\chi}$. This represents a nonsingular 
version of the string since the scalar Goldstone boson, $\chi$, is
canonically conjugate to the Kalb-Ramond two-form, $B_{\mu\nu}$, of the
string (or the three-form $C_{\mu\nu\lambda}$ of the membrane) in four
(five) dimensions.
Suppose we consider this U(1) theory in five dimensions in the background
of the KK monopole:
\be\label{KKMmet}
ds^2 = dt^2 - \left (1 + {E\over r} \right ) d{\bf r}^2
- \left (1 + {E\over r} \right )^{-1} \left [ 
dx^5 + E (1 - \cos\theta) d\varphi \right ]^2
\ee
where $d{\bf r}^2$ represents \real$^3$ in spherical polar coordinates
$(r,\theta,\varphi)$.
In order to obtain a vortex ending on the KK monopole, we must 
choose the phase, $\chi$, to be nonsingular along one of the polar axes.
It is not difficult to see that the choice $\chi = x^5/2E$ satisfies
this requirement. Along the North polar axis, $\Phi$ is regular without
requiring $X=0$, whereas along the South polar axis the metric (\ref{KKMmet})
has a coordinate singularity, and we must transform
to coordinates which are nonsingular in the southern hemisphere: 
${\tilde x}^5 = x^5 + 2E\varphi$, which is the analogue of the Wu-Yang gauge
patching for the Dirac monopole \cite{WY}. 
Hence $\chi = {\tilde x}^5/2E - \varphi$, 
in the vicinity of this axis,
which corresponds to a vortex with winding number $-1$. 

Having decided there is a sensible choice for the phase of the scalar 
field, one might reasonably ask whether this corresponds to a genuine
solution for the vortex. This necessitates solving the $X$-equation
\bea
{\left [ r^2 X_{,r} \right ]_{,r} \over r^2} &+&
{\left [ \sin\theta X_{,\theta} \right ]_{,\theta} \over r^2 \sin\theta}
= {\half} \left ( 1 + {E\over r} \right ) X (X^2-1) \nonumber \\
&+& {X\over4} \left [ {1\over E^2} \left ( 1 + {E\over r} \right )^2 +
{\tan^2 ({\theta\over 2}) \over r^2} \right ] = 0
\eea
where we have chosen to set the vortex width $1/\sqrt{\lambda}\eta=1$.
Fortunately, in the limit $E\gg1$, this can be approximately solved 
analytically, using an analogue of the thin-string approximation used
in \cite{AGK}\ to obtain solutions for vortices in a black hole 
background, by noting that if we set $X = X(R) = X(2\sqrt{r(r+E)}\cos(
{\theta\over2}))$, then the $X$-equation reduces to 
\bea
&-&X''\left [ 1 + {R^2 (3r+2E)\over 4(r+E)^3} \right ]
- {X'\over R} \left [ 1 + {R^2 (5r+6E)\over8(r+E)^3} \right ]
\nonumber \\
&+& {X^2\over R^2} \left [ 1 + {R^2(r+2E) \over 4E^2(r+E)}
\right ] + {\half}X(X^2-1) = 0
\eea
which is simply the equation for the scalar field of a global vortex in
flat space with O($E^{-2}$) corrections. 
Note however, that unlike the flat space global vortex, these corrections mean
that this vortex does not tend asymptotically to the vacuum, but rather 
$X\to 1 - {1\over E^2}$. Obviously this is just an analytic approximation,
and one can integrate the equations numerically, two illustrations of which 
are shown in figure 1. The plots represent the contours of the $X$-field,
($X=0.1,0.2....$) and show how the vortex spreads as $E$ is reduced. Below
$E\simeq 1.7$ the $X=0.9$ contour ceases to exist, and the ripple is a 
manifestation of its marginal nature.
 
\FIGURE{
\includegraphics[height=5cm]{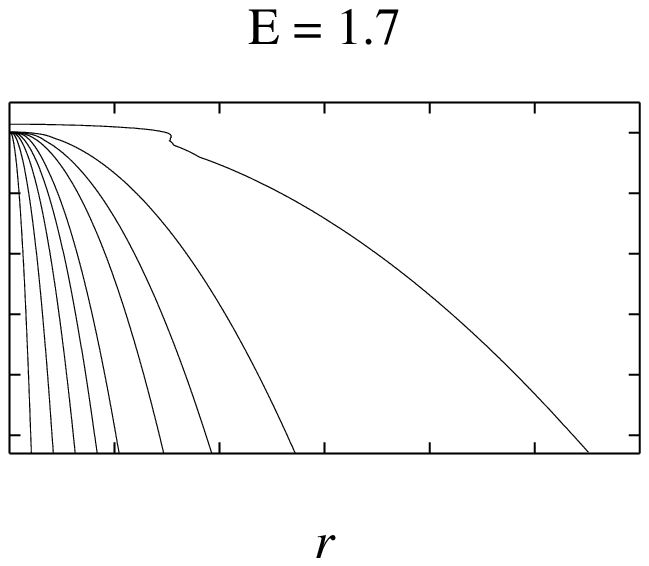}
\includegraphics[height=5cm]{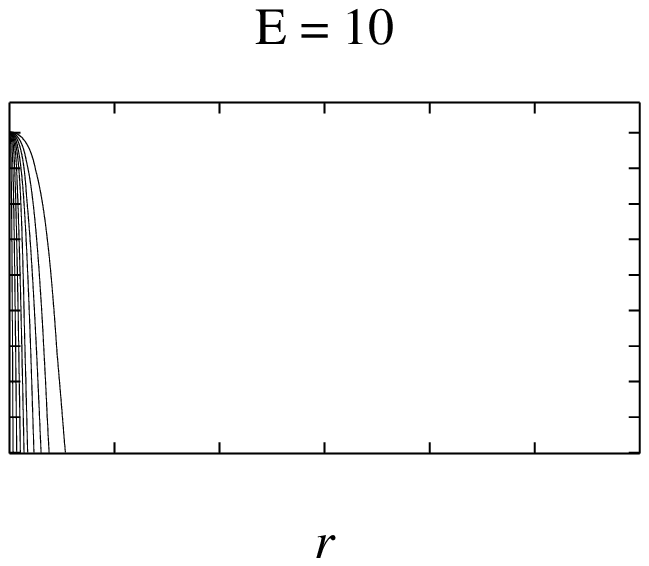}
\caption{$X$-contour lines in the monopole background. The plots are in
``real space'' with the $\varphi$ direction suppressed. The ripple in the
$X=0.9$ contour on the first plot occurs because that contour is marginal
and about to disappear as $E$ drops in value.} 
\label{fig:end}
}

Having used the global vortex as an illustration, one can now read off the
form of the superstring, since in five dimensions a
3-form field is dual to a scalar field, therefore we simply dualise
the solution for $\chi$ in the global vortex to find the four form 
field strength appropriate to the membrane:
\bea
&&F_4 = Q \bigl [ E\sin\theta \left ( 1 + {r\over E} \right )
^2 dt\wedge dr \wedge d\theta\wedge d\varphi \nonumber \\
&+& \tan({\textstyle{\theta\over2}}) dt\wedge dr \wedge d\theta \wedge
[ dx^5 + E(1-\cos\theta)d\varphi] \bigr ]
\eea
When reduced to four spacetime dimensions, this gives rise to the fields
\bea
H &=& \tan ({\textstyle{\theta\over2}}) dt \wedge dr \wedge d\theta \\
{\tilde F}_4 &=& E\sin\theta \left ( 1 + {r\over E} \right )^2 
dt\wedge dr \wedge d\theta\wedge d\varphi
\eea
where $H = dB$ is the `NSNS' two form field strength, and ${\tilde F}_4 
= F_4 + A \wedge H$ is the `RR' four form (with $A$ the RR 1-form).

Notice how $|{\tilde F}_4^2|$ actually tends to a constant at infinity. 
The reason for this
is that in order for the fields to wrap around the internal dimension in such
a way that they close off at the north pole in a regular fashion, yet
still correspond to a string at the south pole, they must wind
around the internal direction, the size of which tends to a 
constant at infinity, therefore the part of the field which winds around
that direction will contribute a constant amount. 
Alternatively, from the four-dimensional point of view, the interaction
of the H-field generated by the string with the A-field coming from the
KK reduction via the Chern Simons term causes a long range effect if we wish
the fields to conspire to make the configuration regular on the north
pole, but have a source on the south pole. In other words, bulk terms 
matter for classical superstrings.
 
In each case the superstring and the global vortex can
be painted on to the KK monopole background in order to give a configuration
corresponding to a terminating string, however the field configuration has 
a nonvanishing local energy density at infinity. This will have
significant implications for the gravitational back reaction of such 
configurations which we will now discuss. 

\section{Gravitational back reaction}

Once we include the gravitational back reaction of the string, we
can no longer assume that the background will be similar to the KK-monopole.
Indeed, there are two very good reasons for supposing that the 
spacetime will be radically different from the monopole, and therefore
that a conventional linearized analysis will not apply. The first
is the presence of these bulk terms -- the long range fields that
are present in order to allow the string to end. The other reason 
is the asymmetry of the configuration. Usually, when one linearizes
around a background configuration one assumes that when the source
is switched on, the spacetime is perturbed in some localised sense.
Here however, we are adding a string along one polar axis of the monopole.
In the case of a standard local cosmic string ending on a black hole,
the solution is altered from the static Schwarzschild solution to
the C-metric \cite{KW}, which although static, contains acceleration horizons
at large radius. In both ways of thinking, the addition of the 
string has had a long range effect.

In order to investigate the gravitational back reaction, instead of
performing a perturbation analysis, we will consider a more general metric
\be\label{kkred}
ds^2 = e^{2\phi\over\sqrt{3}}g_{\mu\nu} dx^\mu dx^\nu - 
e^{-4\phi\over\sqrt{3}}[dx^5 + A_\mu dx^\mu]^2
\ee
Of course, this metric will not be entirely general, the four-dimensional
component will be axisymmetric and static, and the electromagnetic 
potential will be appropriate to a magnetic solution,
$A_\mu = A \partial_\mu \varphi$, such that $A=0$ along the north axis
of symmetry, which will, as before, be the axis pointing away from 
the terminating string. 

To look for a membrane solution in the five-dimensional
metric which is nonsingular
on this axis, we try $F_4=\star 2Q\ dx^5$. Substituting 
the metric (\ref{kkred}) in the five dimensional action 
and integrating out over $x^5$ yields the four-dimensional action
\be\label{action}
S = \int d^4 x \sqrt{g} \bigl [ -R + 2 (\nabla \phi)^2 - {\quarter}
e^{-2\sqrt{3}\phi} F^2 
+ 2Q^2A_{\mu}^2 - 2Q^2e^{2\sqrt{3}\phi}
\bigr ]
\ee
where the factors of $e^{2\phi/\sqrt{3}}$ in (\ref{kkred}) have been 
chosen to put the four dimensional action in the so-called Einstein
frame, in which the gravitational part of the action appears
in the Einstein-Hilbert form. 
In addition to the usual KK terms, this action also contains
a mass term for the KK U(1)-field, and a `cosmological' Liouville potential
for the dilaton.  This action (\ref{action}),
in the absence of the $Q^2 A_\mu^2$ term, has been analysed
with the result that there is a ``no go'' theorem
for spherically symmetric black holes \cite{PW} -- there are no
spherically symmetric black hole solutions which are 
asymptotically flat or de-Sitter.
However, this action and our situation has two crucial differences,
the presence of the mass term for the gauge field, and the lack of
spherical symmetry -- a vortex terminating on a black hole is 
manifestly not spherically symmetric. 

An obvious candidate for a non-spherically symmetric metric is 
of course the dilatonic C-metric of Dowker et.\ al.\ \cite{DGKT}. 
This consists of a black hole under uniform acceleration generated 
by a conical deficit extending towards infinity and is a solution for
the action (\ref{action}) above with $Q^2=0$. 
The key features of the geometry are this conical singularity and an
acceleration horizon generated by the acceleration of the black hole.
For $RA\ll1$, the conical singularity has a deficit
angle of order $RA$, and the acceleration horizon is at a radius of order
$A^{-1}$. Meanwhile, the effects of the Liouville potential will become
relevant at a scale of order $Q^{-1}$, therefore, the dilatonic C-metric 
can only be appropriate as an asymptotic solution if the 
acceleration horizon occurs well before the Liouville and electromagnetic
mass terms are relevant: \ie\ if $A^{-1} \ll Q^{-1}$.

In order to explore this question in greater detail, consider the global vortex.
In Einstein gravity, the global vortex spacetime is
slightly subtle, since in order to have a regular spacetime one
strictly needs to introduce a time dependence or negative
cosmological constant \cite{RG}. Nonetheless, in order to explore the
relative importance of the various terms we will look for a global
vortex solution by maintaining the form of the
Goldstone field $\chi = x^5/R$ (where $2\pi R$ is the periodicity
of $x^5$), then integrating out over $x^5$ gives the four-dimensional
effective action
\bea\label{efflv}
S_{eff} = \int\sqrt{g} \Biggl [ &-&R + 2(\nabla\phi)^2 
- {\quarter} e^{-2\sqrt{3}\phi} F^2 - \eta^2 {X^2\over R^2} 
e^{2\sqrt{3}\phi} \nonumber \\
&+& \eta^2 \left ( (\nabla X)^2 + {X^2\over R^2} A_\mu^2 
-{\lambda \eta^2 \over4}e^{2\phi/\sqrt{3}}(X^2-1)^2 \right )\Biggr]
\eea
The first three terms are the standard KK gravity terms. The last
three terms (grouped together with the factor of $\eta^2$)
come from the kinetic and potential terms of the $\Phi$ field.
However, the kinetic term of the Goldstone $\chi$-field gives
two contributions (as seen in the $Q^2$ terms of (\ref{action})),
one of which is the gauge field term and the other the Liouville term
of $X^2 e^{2\sqrt{3}\phi}$. The reason for putting this term seperately
is to emphasize the grouping of the $\eta^2$-terms. These are very close
to the lagrangian terms for a local U(1)-vortex, indeed,
if we identify $1/R$ with $e$, then apart from the Liouville
term, this is precisely the lagrangian
of a dilatonic {\it local} vortex, with Bogomolnyi parameter
$\beta = 1/2e^2\eta^2 = R^2/2\eta^2$.  (Note that this is reminiscent
of the KK vortices of Dvali, Kogan and Shifman \cite{DKS},
however they {\it broke} the U(1) $\partial_5$ Killing symmetry
with a braneworld, whereas we have not.)
The gravitational interactions
of such vortices were studied in \cite{GS}, where it was shown
that they could be used to smooth out conical deficits in standard
`vacuum' spacetimes {\it including} those with dilatonic black holes. 

Briefly, in \cite{GS} it was shown that a dilatonic local vortex, like
the standard cosmic string, has a conical deficit in the vicinity of
the core, which smooths out the apex of the cone. However, the main
difference between dilatonic and Einstein strings is that, depending on
how the local vortex couples to the dilaton, it is now possible for
spacetime to be non asymptotically locally flat far from the core, 
with the dilaton providing the curvature. Only those vortices which
couple canonically to the dilaton (\ie\ in the string frame if we have
string gravity, or the KK frame for KK gravity) avoid this fate. In
general in Einstein gravity, it is always possible to use a local cosmic
string to smooth out the conical deficits present in a metric \cite{smooth}, 
for example the C-metric which represents two accelerating black holes
being pulled away from each other by conical deficits extending
to infinity. In dilaton gravity, it was shown in \cite{GS} that a dilaton
vortex threading a black hole would, if noncanonically coupled, add
dilaton charge to the black hole and asymptote the vortex metric. However,
for the dilatonic C-metrics, the conical deficit could {\it only}
be smoothed if the coupling of the vortex to the dilaton was canonical.
Fortunately, the coupling in (\ref{efflv}) is precisely canonical.

Now, the
analytic analysis in \cite{GS} was undertaken supposing $\beta \simeq O(1)$,
whereas we have $\beta = R^2/2\eta^2$, with $R\gg1$ (large KK monopole
mass) and in addition $\eta\ll1$ (as $\eta^2$ really stands for $8\pi G\eta^2$
in vortex units, and we are expanding around the low string mass limit).
Therefore, we are in a relatively little explored parameter space of
the local vortex (some work was done on large $\beta$ vortices interacting
with non-dilatonic black holes in \cite{BEG}). 
However, by combining results and intuition
from local vortices with black holes, as well as the global vortex itself,
we can in the absence of the `Liouville' term describe the spacetime.

There will be three main regions: $r\ll R$, $R\ll r\ll R\eta^{-1}$,
and $r\gg R\eta^{-1}$, where $r$ is to be understood in a qualitative
sense as representing the distance scale on which one is examining
the spacetime. Roughly speaking these three scales represent the
scale on which spacetime is inherently five-dimensional in nature;
where the solution is four-dimensional in nature but the Liouville 
term is not yet relevant;
and finally the long range effects of all the terms in the action.

The immediate effect of the vortex core is to provide a snub-nosed conical
deficit. This effect is strongly localized around the core of the vortex,
and therefore ought not to be particularly affected by the Hopf fibration
of the magnetic black hole. There will therefore be a region around 
the south polar axis which is a conical deficit in the $\varphi$ polar angle.
At intermediate scales, we are well outside the vortex core, and
in a four-dimensional r\'egime, yet still well below the scale 
at which the Liouville and mass terms are relevant. Since the vortex
couples in the KK frame, we expect from \cite{GS} that the dilatonic
metric of \cite{DGKT} will describe the metric at this scale (with the
vortex smoothing out the core of the conical deficit). Finally, at
large scales the electromagnetic flux confines, and in the absence of
the Liouville term we would simply have a vacuum C-metric.

Such is the spacetime in the absence of the Liouville term. Now let us
consider inclusion of this term, which becomes relevant at $O(R\eta^{-1})$.
First of all,
if this scale is outside the acceleration horizon of the C-metric, then
it is clearly irrelevant. The acceleration horizon of the C-metric is at
$r\sim O(A^{-1})$, where $RA\sim \eta^2$ is the deficit angle of the
C-metric. Thus $A^{-1} = O(R\eta^{-2}) \gg R\eta^{-1}$ for $\eta\ll1$.
Therefore the Liouville scale is well within the acceleration horizon --
a conclusion we expect to be true for the string/membrane as well.
For the global vortex, note that it is precisely this Liouville
term which is inherently a global vortex term. If we were simply
looking for a five dimensional global vortex membrane, then we could
always consider reducing over the angular variable which represented
the winding of the Goldstone field. This would give rise to precisely
that Liouville term. For a simple cylindrically symmetric vortex in
the absence of the electromagnetic field,
it is this term which is responsible for the singularity in the static metric
for the vortex \cite{CK}. However, to render the global vortex
nonsingular, one can add a negative cosmological constant, or
worldbrane curvature \cite{RG}, the latter of which
causes a compactification of spacetime which becomes a small 
deformation of the de Sitter hyperboloid \cite{GS2}.

Applying these results to the case at hand, one would therefore
expect that the unchecked Liouville term could well render 
the acceleration horizon of the C-metric singular, however, one
can remove this singularity by adding a small negative cosmological
constant, or time-dependence to the metric. Since adding time-dependence
means that the metric will now depend on three variables, a rather
difficult problem, we will simply assume that a tiny negative cosmological
constant has been added to cancel the Liouville term in an analogous
way to the self-gravitating global vortex, and we are left with a vacuum
C-metric with a minutely small acceleration.

\section{Discussion}

So, overall the geometry has three different approximate descriptions:
First, the near-field, or core, which for the global string
is a snub-nosed cone -- for the superstring, this will
simply be its local near-core metric. Second there will be a mid-field
approximation  in which the geometry and dilaton will have the form
of the dilatonic C-metric. Finally, on the large scale, the mass
term for the electromagnetic field will cause that flux to confine 
as well, and (with the proviso of a checking-term in either
the action or the intrinsic worldbrane geometry)
we will have a standard C-metric. 

We can ask how these conclusions are altered if we try to
explore a different range of parameter space, either by lowering
the compactification radius $R$, or increasing the gravitational
strength of the global vortex $\eta^2$. The plots in figure \ref{fig:end}
show that for the global vortex at least, we can lower $R$ to a similar
order of magnitude as the vortex width before we run into trouble.
Similarly, analytic arguments on the existence of vortex solutions
in \cite{GS2} show that we can raise $\eta^2$ to $O(1)$ without 
destroying the general conclusions discussed here. We therefore expect
that these qualitative results hold true even for strings ending on 
monopoles with a similar mass. 

Clearly this is only a toy model, and cannot directly be used
to draw any conclusions for the ten-dimensional problem, however,
there are some interesting features which crop up here that might have some
analogue in the higher dimensional case. One is the presence of the
bulk terms giving the cosmological term in the action. This was a feature
of the finite size of the compactification radius in five-dimensions.
While we might expect this effect to be ameliorated in higher
dimensions, in that it may not lead to a cosmological term, we
might expect that the energy density of the various RR and NSNS
fields will not fall-off as rapidly as for the superstring itself.
The other curious feature we have seen is the mass term for the RR-field, 
which causes confinement of the magnetic flux. Again, this effect could
be lessened, but it does give an interesting potential picture for the
distortion of the flux around the intersection point. In fact, if one
plots the magnetic flux of the near horizon solution for the string ending 
on the monopole one does see some evidence of this.

The other main way in which a higher dimensional problem will differ
is in the number of degrees of freedom of the solution. By working
in only four dimensions, the problem reduced to an effectively 
two-dimensional question: the distance from the monopole core, and the
angle from the string. We can view any classical supergravity problem
as a dimensional reduction over the `inessential' coordinates (\ie\ those
that represent symmetries of the metric) down to the space on which
the metric depends. For two variables we can always express this space
as conformally flat. However, as soon as the string ends on an extended brane,
there are three physical variables - the distance along the string
from the endpoint, the distance along the brane from where the string
touches, and finally, the mutually orthogonal distance from the system.
Unfortunately, we cannot write a three-dimensional problem in
a conformally flat, or even necessarily diagonal fashion. This shows
up in the perturbative analysis of \cite{plot}. Since our
universe is however four dimensional up to fairly high energies,
it is reasonable to make a four-dimensional approximation to the problem.

However, it is amusing that the act of ending the string on the black hole
causes its flux to confine, and therefore removes all
evidence of its charge from the asymptotic observer, and leaves
it with only a `label' attached to the end of the string.
Whereas Nielsen and Olesen \cite{NO} originally used the U(1) abelian
Higgs model to construct a physical realisation of the Nambu-Goto string;
by considering a field theoretic realisation of an open
superstring, a flux confinement mechanism switches on,
and the string ends up as a truly confined flux tube with (abelian --
as we have a unit charged monopole) `quarks' at each end.

\section*{\bf Acknowledgements.}
 
I would like to thank Peter Bowcock, Simon Ross and Douglas Smith for 
useful discussions, and particularly Filipe Bonjour for assistance with the 
numerical work. This work was supported by the Royal Society.

\end {document}